\documentclass[pre,aps,twocolumn,superscriptaddress,floatfix]{revtex4-1}
\usepackage[english]{babel}

\usepackage{dcolumn}
\usepackage{amssymb}
\usepackage{amsmath}
\usepackage{bm}
\usepackage{color}
\usepackage{graphicx}

\usepackage[usenames,dvipsnames]{xcolor}

\usepackage{color}
\definecolor{darkblue}{rgb}{0,0,0.6}
\definecolor{darkred}{rgb}{0.6,0,0}

\usepackage{hyperref}
\hypersetup{
colorlinks=true,			
urlcolor=darkblue,
citecolor=darkblue,		
linkcolor=darkred,	
}

\begin{document}

\title{Elastic interfaces on disordered substrates: From mean-field depinning to yielding}

\author{E. E. Ferrero} 
\affiliation{Instituto de Nanociencia y Nanotecnolog\'{\i}a, CNEA--CONICET, 
Centro At\'omico Bariloche, (R8402AGP) San Carlos de Bariloche, R\'{\i}o Negro, Argentina.}

\author{E. A. Jagla} 
\affiliation{Centro At\'omico Bariloche, Instituto Balseiro, 
Comisi\'on Nacional de Energ\'ia At\'omica, CNEA, 
CONICET, UNCUYO,\\
Av.~E.~Bustillo 9500
(R8402AGP) San Carlos de Bariloche 
R\'io Negro, Argentina}

\begin{abstract} 
We consider a model of an elastic manifold driven on a disordered energy landscape,
with generalized long range elasticity.
Varying the form of the elastic kernel by progressively 
allowing for the existence of zero-modes, the model interpolates smoothly between 
mean field depinning and finite dimensional yielding.
We find that the critical exponents of the model change smoothly in this process. 
Also, we show that in all cases the Herschel-Buckley
exponent of the flowcurve depends on the analytical form of the microscopic
pinning potential.
This is a compelling indication that within the present elastoplastic description yielding in finite
dimension $d\geq 2$ is a mean-field transition.
\end{abstract}

\maketitle


Statistical physics is built on analogies.
The comparison of typically complex problems with a small
number of simpler ones for which an exact solution is known
is the first step in almost every argumentative construction.
For instance, the so-called out-of-equilibrium phase transitions
have been discussed in the mirror of equilibrium phenomena.
The problem of depinning of an elastic manifold moving
on a disordered landscape has been rationalized by analogy with the
theory of critical phenomena~\cite{FisherPR1998,KardarPR1998}
and studied for over 30 years already.
Once this problem has been reasonably understood, it serves
in turn as the base model for a new analogy step.
In this sense, depinning has shaped the theoretical endeavors
in the understanding of the yielding transition of amorphous solids
under deformation, that received full attention of the statistical
physics community only recently.
The problem with analogies is that, some times, they may prevent to
see the big picture.

Members of the family of sand-pile problems, both depinning and yielding
are paradigmatic examples of driven transitions and are intuitively very alike.
Depinning is related to the movement of an elastic manifold in the
presence of a quenched disordered potential, under the action of
an external driving force.
Yielding pertains to the flow of an amorphous solid upon the
application of an external driving stress or deformation.
In both cases, if the driving force is weak (and the possibility of thermal
activation is excluded), the system remains in a frozen configuration;
however, if a critical threshold is exceeded, the system reaches a
dynamical state with a non-zero average velocity (depinning picture)
or strain rate (yielding picture). 
The critical threshold defines the transition. 
In depinning the velocity-force characteristics of the system shows
singular behavior at a critical force $f_c$. 
While $v=0$ for $f<f_c$, it behaves as
$v\sim (f-f_c)^\beta$ when $f$ increases above $f_c$, with $\beta$ a
well defined number known as the flow exponent.
In yielding the transition is characterized by the critical behavior of
the strain rate $\dot \gamma$, which is zero when the stress
$\sigma$ is  below a critical value $\sigma_c$, and becomes
$\dot \gamma\sim(\sigma-\sigma_c)^\beta$ when $\sigma>\sigma_c$.
The value of $\beta$ is referred again as the flow exponent. 
Its inverse $n \equiv 1/\beta$ is known as the Herschel-Buckley exponent. 

The depinning transition finds a continuous model approach in the Quenched
Edwards-Wilkinson equation, which allows for analytical treatment using
Functional Renormalization Group (FRG) analysis.
For yielding, elasto-plastic models (EPMs) built at a coarse grained 
scale~\cite{NicolasRMP2018} provide a similar
description~\cite{TalamaliPRE2011,LinPNAS2014,BudrikisNC2015}.
Nevertheless, despite the analogous construction, an FRG treatment of EPMs
has found limitations
and analytical support for a theory of yielding is only provided so-far
by mean-field variants~\cite{HebraudPRL1998, AgoritsasEPJE2015, JaglaPRE2015,
AgoritsasSM2016,LinPRX2016, lin2017some, AguirrePRE2018, JaglaJSTAT2018}.
In EPMs the instantaneous values of stress and plastic strain are
evolved consistently.
Under a condition of uniform load, the stress increases uniformly.
When the stress exceeds locally a threshold value, the local plastic strain
increases at that patch, causing a reduction of the local stress and also
a perturbation of the stress in every other point in the system, following
the action of elastic interactions.
The form of this interactions is the one prescribed by the Eshelby
propagator~\cite{Eshelby1957,Picard2004} of continuum mechanics;
which has a $\sim 1/r^d$ spatial decay and thus it is a
long-range interaction. 
%
Also, it has alternating signs depending of the direction, with a quadrupolar
symmetry.
This anisotropy is a curse for the FRG approach 
and the responsible for special avalanche correlations in the form of slip
lines (or planes), that greatly determine the differences among the plastic
yielding transition and its elastic depinning counterpart.

%
In the present paper we show that mean-field depinning and yielding
transition in finite dimensions can be considered to be special cases of
a generalized mean-field problem and, therefore, described within the same framework.
Simply considering an elastic kernel as sum of two contributions $G^{\tt MFD}$
and $G^{\tt Y}$, corresponding respectively to a constant value propagator
(in Fourier space) and the Eshelby propagator, we are able to \textit{smoothly}
interpolate ($\varepsilon G^{\tt MFD} + (1-\varepsilon) G^{\tt Y}$ with
$0\geq \varepsilon\geq 1$) between mean-field depinning and yielding.
In particular, we observe a smooth transition in the values of the critical exponents
between the two limiting cases.
Thus our work suggests an alternative view for the theoretical
tackling of the yielding transition, 
interpreting it as a particular case of a general mean field problem that includes
also the fully-connected mean-field depinning. 

\subsection*{A general model for MF-depinning and yielding}

The general model that allows to describe MF-depinning and yielding on the
same footing is constructed in the following way.
The variable of the model is a scalar field $e_i$ defined on the sites $i$
of a $d$ dimensional ordered lattice. 
For depinning $e_i$ represents the interface position at site $i$,
whereas for yielding $e_i$ is the strain of an elemental volume
of the system at site $i$. 
The dynamics is described by overdamped equations of motion of the form
\begin{equation}
\eta\frac{de_i}{dt}=f_i(e_i)+\sum_j G_{ij}e_j+\sigma
\label{model}
\end{equation}
In the case of depinning, the terms $f_i(e_i)\equiv-dV_i/de_i$ represent
the force exerted by the external pinning potential $V_i$,
whereas for yielding they describe the local thresholding behavior
of a small piece of the amorphous material under deformation.
In both cases the form of the functions $V_i$ are of the same kind:
they have minima at different $e_i$ positions representing local
equilibrium states.
In Eq.(\ref{model}) $G_{ij}$ represents the elastic interaction between
$e$ values at different points. 
We restrict to cases in which this interaction preserves the homogeneity
of the system, then $G_{ij}$ depends only on the difference between the
(vector) positions $i$ and $j$. Also, 
$G_{ij}=G_{ji}$ is assumed.
Elastic forces should be balanced in the system, therefore $\sum_i G_{ij}=0$ must be also satisfied.
This still leaves us with a lot of freedom in the choice of a general form for $G_{ij}$.
Nevertheless, an important additional constraint must be fulfilled:
in the absence of local forces ($f_i\equiv 0$) the flat configuration of the interface
$e_i={\tt cst}$ must be stable.
This condition becomes more transparent in Fourier space, where Eq.(\ref{model}) reads for ${\bf q}\ne 0$

\begin{equation}
\frac{de_{\bf q}}{dt}=f(e)|_{\bf q}+G_{\bf q}e_{\bf q}
\label{modeloq}
\end{equation}
The stability condition is then $G_{\bf q}\le 0$.

In the following we will mainly discuss the interaction kernel
in Fourier space.
One can consider ``generalized mean field models" defined as 
cases in which the $G_{\bf q}$ is zeroth order homogeneous in $|q|$.
These kernels produce a function $G_{ij}$ that is either independent
of distance or decaying with $r_{ij}$ as $r_{ij}^{-d}$.
In any of the two cases, the effect of a single site onto another site
is negligible compared to the combined effect of all other sites in the lattice.
Therefore, the dynamics of a given site can be solved by considering the
existence of a (fluctuating) prescribed external field (see \cite{AguirrePRE2018}).
In particular, the forms of $G_{\bf q}$ for mean-field depinning and yielding
satisfy the prescription just mentioned.
For mean field depinning $G^{\tt MFD}_{\bf q}= -1$ for ${\bf q}\ne 0$, 
whereas for yielding $G_{\bf q}$ is the Eshelby propagator that in $2D$
can be written as (${\bf q}\ne 0$)
\begin{equation}
G^{Y}_{\bf q}=-\frac{(q_x^2-q_y^2)^2}{(q_x^2+q_y^2)^2} 
\end{equation}
In both cases $G_{{\bf q}=0}$ is taken as zero in a stress conserved dynamics,
as it follows from the condition $\sum_i G_{ij}=0$.
The uniform mode in Eq.(\ref{model}) is thus directly found from (we set $\eta\equiv 1$ for the rest of the paper)
\begin{equation}
\dot\gamma\equiv \frac{d\overline{e_i}}{dt}=\overline{f_i(e_i)}+\sigma
\label{gammadot}
\end{equation} 
that defines the global strain rate $\dot\gamma$.
The fact that both $G^{\tt MFD}$ and $G^{\tt Y}$ share the property of being $\mathcal{O}(q^0)$,
allows us to believe that mean-field depinning and yielding may share many common features.

\begin{figure}[!bt]
\includegraphics[width=0.9\columnwidth,clip=true]{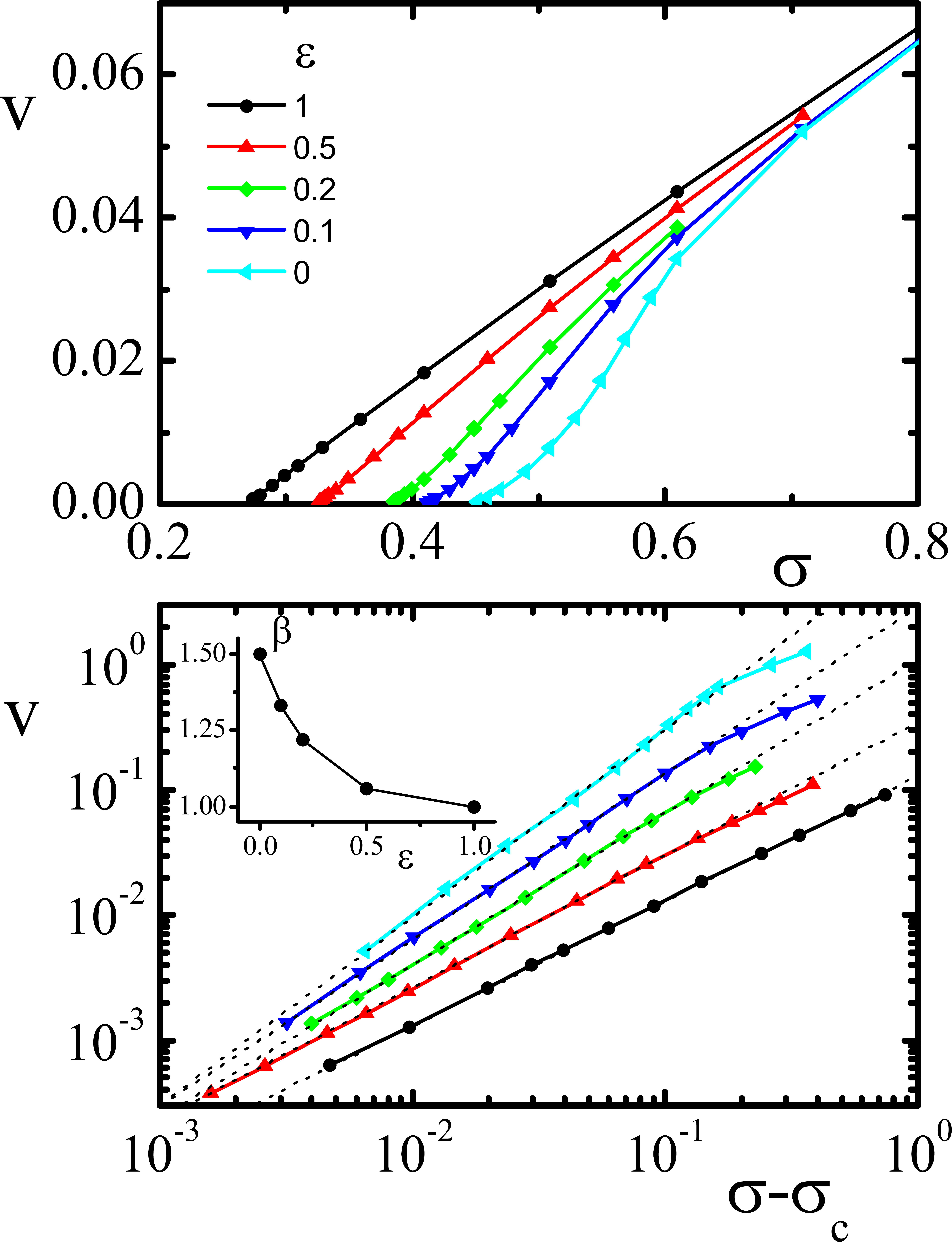}
\caption{
Flow curves in a system of size $N=512^2$,
interpolating between the mean field depinning case ($\varepsilon=1$)
and the yielding case ($\varepsilon=0$) in linear (a) and logarithmic scale (b).
The inset in panel (b) shows the values of $\beta$ determined as the slope of the straight lines,
for $\sigma\to \sigma_c$ (in panel (b) the curves were progressively displaced vertically,
to facilitate visualization).
\label{beta}
}
\end{figure}

With respect to the properties of the disorder term $f_i(e_i)$,
we restrict to the case of locally correlated potentials, where
$\langle f_i(z)f_i(z+\Delta)\rangle$ decays to zero sufficiently
fast with $\Delta$.
Also, we consider the disorder site by site to be totally uncorrelated, namely $\langle f_i(z)f_j(z)\rangle=0$ for $i\ne j$.
With these correlation properties, Renormalization Group theory teaches that the detailed form
of  $f_i(z)$ should be irrelevant when determining the critical properties of the transition,
as long as the elastic interaction $G_{ij}$ decays sufficiently fast in space as a
function of $r\equiv |r_i-r_j|$,
concretely, if $G_{ij}\sim r^{-\alpha}$ with $\alpha>d$.
In cases in which $\alpha \le d$ (there included mean-field depinning and yielding) this result does not apply,
and different values are obtained for the dynamical exponents when considering
``cuspy" or ``smooth" potentials~\cite{FisherPR1998,KoltonPRE2018}.
In the simulations presented below, we mainly focus in the case of a cuspy form of the pinning potential,
taking $V_i$ as composed by a concatenation of parabolic pieces.
A brief consideration of the smooth potential case is included at the end.

\subsection*{Results}

We present simulations in $2D$ using a kernel
\begin{equation}
G_{\bf q}\equiv (1-\varepsilon)G_{\bf q}^{\tt Y} + \varepsilon G_{\bf q}^{\tt MFD}
\label{eq:combinedG}
\end{equation}
that interpolates between the mean field depinning case (for $\varepsilon=1$) and
the yielding case (for $\varepsilon=0$).
The stress-controlled and quasistatic strain-controlled protocols to determine
the flowcurves and the avalanche statistics are described in Appendix~\ref{ap:protocol} in detail.

Both for an elastic interface undergoing depinning and for an amorphous solid at
the onset of yielding, a singular behavior of $\dot{\gamma}$ (strain-rate or velocity)
is expected at $\sigma_c$.
%
Fig.~\ref{beta} shows the flow curves for different values of $\varepsilon$.
By plotting the data in logarithmic scale close to $(\sigma-\sigma_c)$ 
\footnote{$\sigma_c$ depends on the value of $\varepsilon$ and must be fitted for each curve},
a clear power-law behavior allows to determine the flow exponent $\beta$. 
Going from $\varepsilon=1$ ($\tt MFD$) to $\varepsilon=0$ ($\tt Y$) we observe that the $\beta$
exponent moves from $\beta=1$ to $\beta\simeq 1.5$.
Very importantly, this variation is {\em smooth} as the inset in Fig.\ref{beta}(b) shows,
indicating the continuous evolution that exists between mean-field depinning and yielding.


We now discuss the avalanche size distribution $P(S)$
associated to the transition.
When $P(S)$ is taken from large collections of avalanches
obtained in a quasi-static simulation,
it is expected to be power-law distributed,
namely $P(S)\sim S^{-\tau} f(S/S_{\tt max})$ with the cutoff
function $f(x)$ behaving as $f_{x\to 0}\to 1$ and
$f_{x\to \infty}\to 0$, and $S_{\tt max}$ depending on the
system size $L$ and the stress non-conserving parameter $\kappa$
used to define the value of $G_{{\bf q}=0}$ in the quasi-static protocol.
\begin{figure}[!tb]
\includegraphics[width=0.9\columnwidth,clip=true]{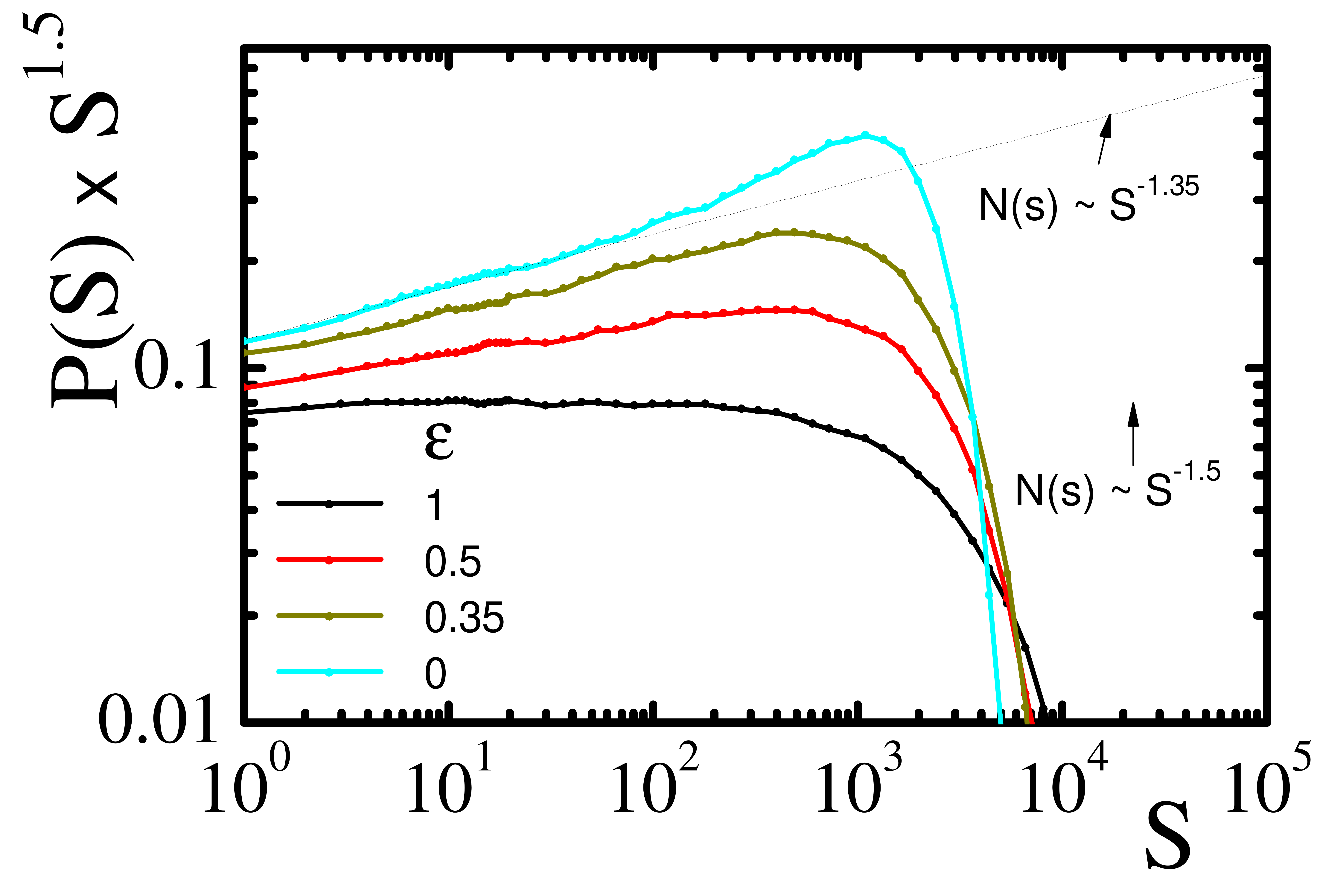}
\caption{Avalanche distribution $P(S)$ interpolating between MF depinning 
($\varepsilon=1$) and yielding ($\varepsilon=0$), note the particular
scaling of the vertical axis, to emphasize differences (curves where also vertically displace, for clarity).
System size is in all cases $N=512^2$.
\label{avalanches}
}
\end{figure}
Avalanche size distributions are shown in Fig.~\ref{avalanches} 
for different $\varepsilon$.
\footnote{For a better comparison, in Fig.~\ref{avalanches} the values of $\kappa$ were
adjusted to obtain curves with a similar cutoff.} 
In depinning mean-field models $\tau=\frac32$, and
in fact, we obtain $\tau=1.5$ when $\varepsilon=1$.
But as we decrease $\varepsilon$ moving towards yielding,
$\tau$ diminishes, becoming $\tau\simeq 1.35$ at $\varepsilon=0$.
Surprisingly, this change is \textit{continuous}; meaning, without scale crossovers
on the observable.
The avalanche size distribution critical exponent is a smooth function of
the parameter $\varepsilon$.

Directly related to the avalanche mean size is the loading stress
needed to trigger avalanches $x_{\tt min}$.
It is known for yielding~\cite{Maloney2004,KarmakarPRE2010,LinPNAS2014}, that its mean
value scales sub-extensively with system volume $N$:
$\overline{x_{\tt min}} \propto N^{-\alpha}$, with $1<\alpha<2$.
%
%
This phenomenological sub-extensiveness in the plastic flow of amorphous solids
under deformation was interpreted~\cite{LinEPL2014,LinPNAS2014} as a consequence
of a peculiar shape for the steady state distribution $P(x)$ of local distances to
threshold $x$ \footnote{In a driving protocol towards the right we define $x_i=e_{Ri}-e_i$ 
according to the notation in Appendix \ref{ap:protocol} }.
If this quantity has the form $P(x)\sim x^\theta$ as $x\to 0$, one can
deduce~\cite{KarmakarPRE2010} that $\overline{x_{\tt min}} \propto N^{-1/(1+\theta)}$.
Then, $\theta=0$ is expected for depinning-like models (where the kernel $G_{ij}$ is non-negative)
and $\theta>0$ for yielding models (where the kernel $G_{ij}$ alternates in sign).

\begin{figure}[!tb]
\includegraphics[width=0.9\columnwidth,clip=true]{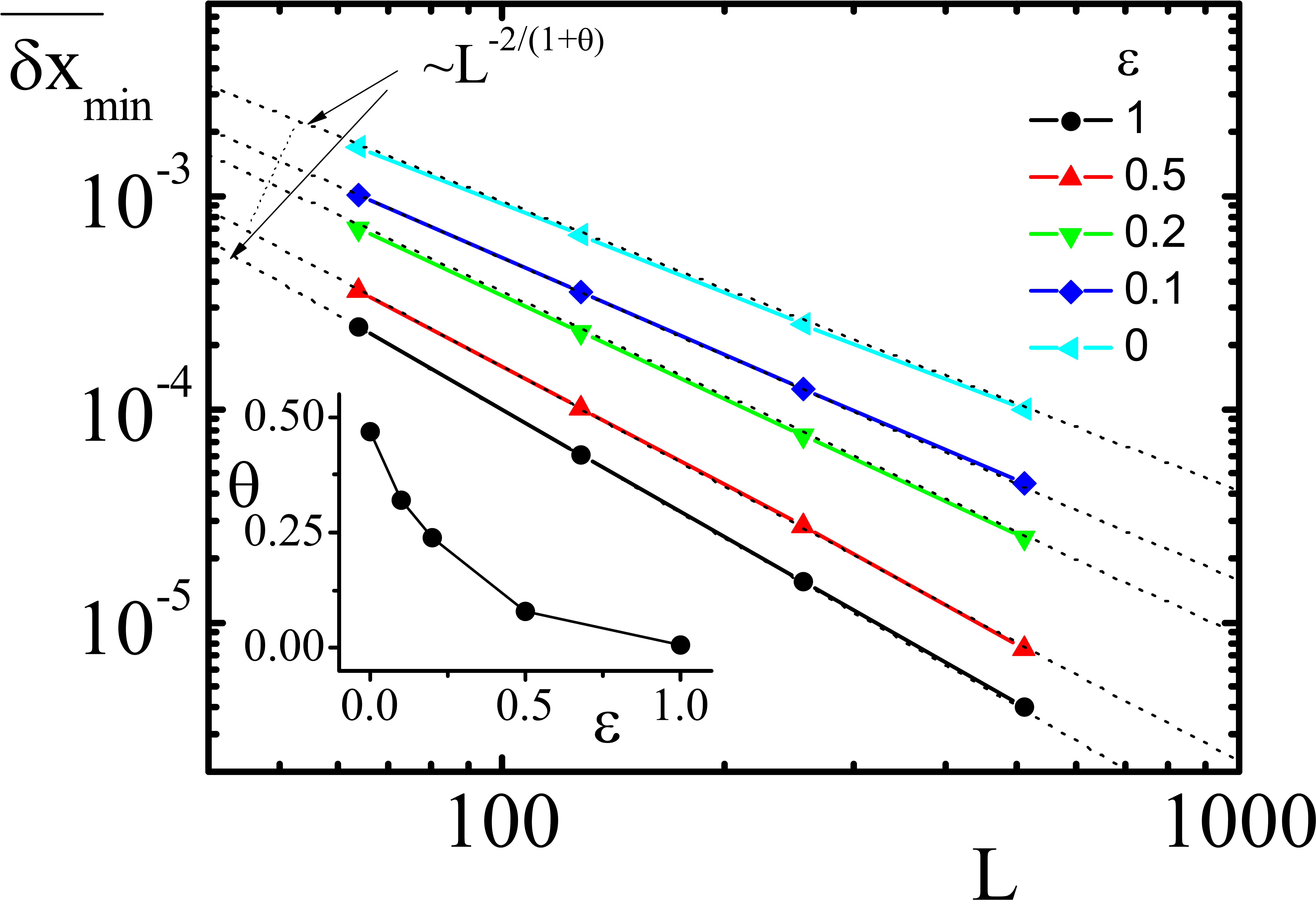}
\caption{
Evolution of $\overline{x_{\tt min}}$ as a function of system size $L$,
for different values of $\varepsilon$. By fitting with straight lines of slope $-d/(1+\theta)$ 
the values of $\theta$ are obtained, and plotted in the inset.
\label{theta2}
}
\end{figure}

Figure~\ref{theta2} shows results for $\overline{x_{\tt min}}$ vs $L$
for systems with different $\varepsilon$.
Power-law fits allow for a precise determination of the exponent values.
Consistently with the expectation, $\theta=0$ for MF-depinning ($\varepsilon=1$)
and a strictly positive value for the yielding case ($\varepsilon=0$).
What is surprising again is that $\theta$ turns out to be a \textit{continuous}
function of the crossover parameter $\varepsilon$, going from $0$ to $\simeq 0.5$
as we move from the MF-depinning limit to the 2$d$-yielding limit,
as displayed in Fig.~\ref{theta2} inset.
This tells us that we are working with a family of similar problems, each of them characterized
by a given degree of sub-extensiveness of the load needed to trigger new avalanches.

{\it Smooth pinning potentials--}
All results presented so far were obtained using a local disorder potential
that has cusps in the transition from one potential well to the next one.
Usually, according to renormalization arguments, this kind of details on
the microscopic potential should not influence the critical properties of
a system.
In particular, the critical exponents of the depinning transition are expected
to be independent on the potential being of the cuspy or smooth type.
Nevertheless, the fully connected mean-field case is an exception
(see discussion in~\cite{KoltonPRE2018}).
There, we know that depinning displays a value $\beta=1$ for cuspy pinning potentials
and a (different) value $\beta=3/2$ for smooth pinning potentials.

The smooth crossover of exponents that we observe between mean-field depinning
and yielding in Fig.\ref{beta}, suggests that we will also find the above described
dichotomy in the flow curve exponent value for the yielding case.
Even more, we can expect to find larger values of $\beta$ using smooth potentials for
any value of the crossover parameter $\varepsilon$.
\begin{figure}
\includegraphics[width=0.9\columnwidth,clip=true]{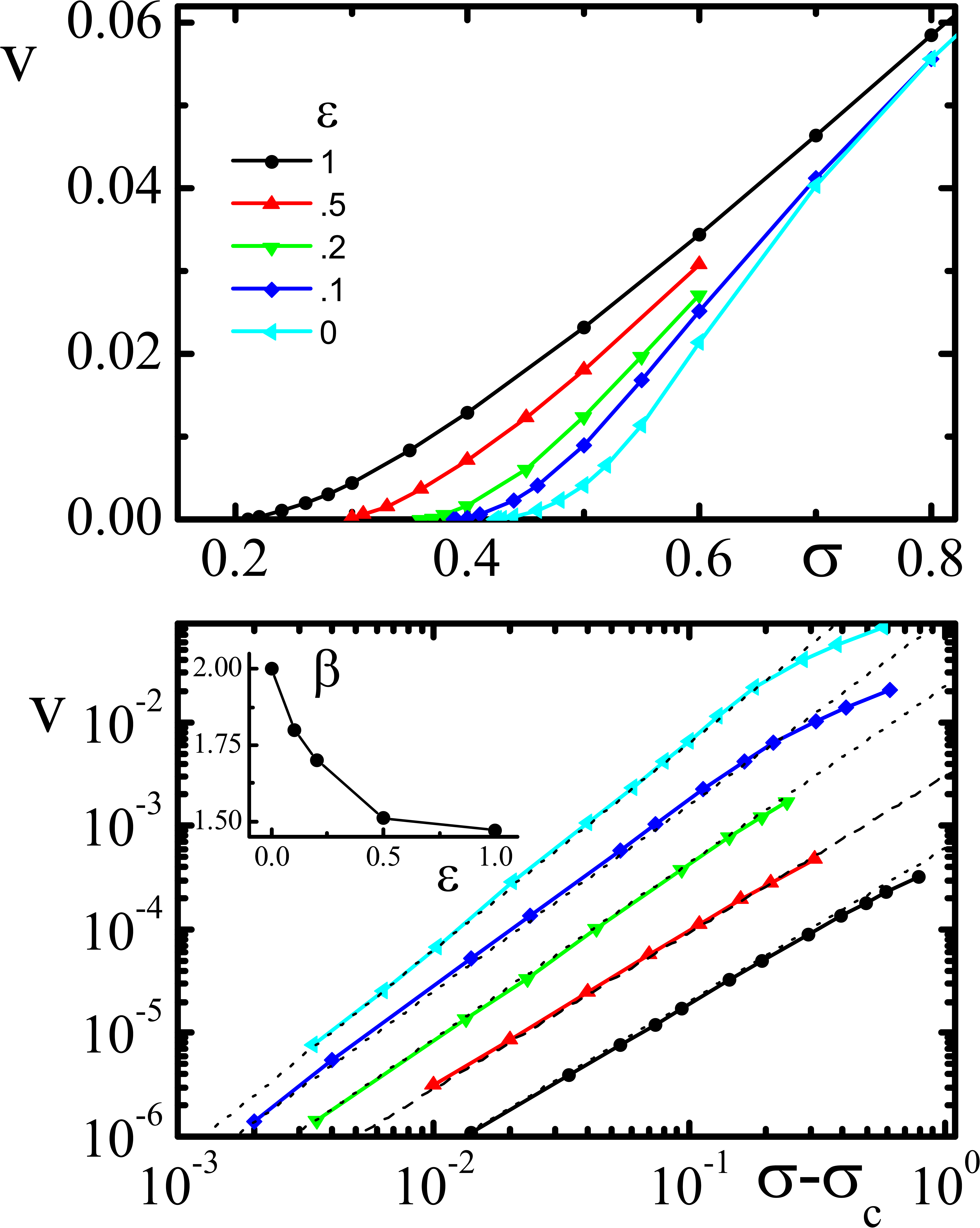}
\caption{
Same as Fig.\ref{beta} but using a smooth pinning potential.
The values obtained for $\beta$ at corresponding values of
$\varepsilon$ are found to be 1/2 larger than those for cuspy potentials. 
\label{beta_smooth}
}
\end{figure}
Results of simulations contained in Fig.~\ref{beta_smooth} confirm this result.
Note that the $\beta$ value for smooth potentials always (i.e., for each $\varepsilon$)
exceeds in $\frac12$ the one for cuspy potentials, in full agreement with recent 
theoretical expectations derived from the Prandtl-Tomlinson model under stochastic driving~\cite{JaglaJSTAT2018}.

{\it Three and larger dimensional cases--}
A scalar Eshelby kernel for $d=3$ in Fourier space (taking a diagonal
component of the non-deviatoric stress tensor) can be written as
\begin{equation}
G_{\bf q}^{\tt 3D} =  \frac{2q_x^2(q_y^2+q_z^2)}{(q_x^2+q_y^2+q_x^2)^2} -1
\end{equation}
One can notice that again $G_{\bf q}\sim {\bf q}^0$,
and this is true in general for $d>1$. 
Therefore, we expect all our numerical observations and conclusions
obtained in $d=2$ to be valid also in $d=3$ and higher dimensions. 
Preliminary simulations in $d=3$ for the cuspy potential give us, for instance,
a smooth crossover between $\beta^{\tt Y}\simeq 1.3$ (not far from other
estimations~\cite{LinPNAS2014}) and $\beta^{\tt MFD}= 1$ continuously moving with $\varepsilon$.
The value of the pseudo-gap exponent $\theta$ also changes continuously with $\varepsilon$.
The reduction of $\beta$ in passing from $2$ to $3$ dimensions can be rationalized as a consequence
of the reduced density of zero modes in the elastic propagator in $3d$ compared to $2d$. 

{\it Why a smooth exponent crossover is surprising?--}
A remarkable point is that the critical exponents (values of $\beta$, $\theta$, and $\tau$ 
in particular) vary smoothly between those of $\tt MFD$ and $\tt Y$.
This situation is not expected in general when studying crossovers between different asymptotic behaviors.
Consider for instance the case of long range depinning.
Choosing a kernel decaying in space as $G_1\sim 1/r^{\alpha_1}$ ($d<\alpha_1<d+2$) a set
of critical exponents is obtained.
For other decaying form of the kernel $G_2\sim 1/r^{\alpha_2}$ the exponents are different.
However, if we combine the two kernels in the form $G=(1-\varepsilon)G_1+\varepsilon G_2$
the system will display the critical behavior corresponding to the lowest value of $\alpha$.
In other words, if two \textit{different} criticalities are mixed together the system
will display at long enough scales the critical exponents corresponding to the longest
range interactions.
In order to have a variation of the critical exponents with $\varepsilon$, the long range
weight of $G_1$ and $G_2$ must be similar.
This non-common case is, in fact, what takes place in our kernel combination; clearly manifested
in the ${\bf q}$ space form of the propagator: both the constant ($\tt MFD$) and the Eshelby ($\tt Y$)
kernels scale as ${\bf q}^0$.
The $\tt MFD$ kernel in Fourier space has a finite constant value, then the value of the
interaction with any other site in real space is of order $\sim 1/L^d$.
In real space, the Eshelby kernel decays as $1/r^d$ and, then, its typical value at distances
of the order of the system size $L$ is also $\sim 1/L^d$.

\subsection*{Conclusions}

We have studied a mesoscopic implementation of a scalar model for a generalized 
elastic manifold on a disordered landscape that is able to describe both depinning
in mean-field ($\tt MFD$) or finite dimensional yielding ($\tt Y$)
by changing the form of an elastic kernel interaction.
The most important result is the observation of a \textit{smooth} transition between
$\tt MFD$  and $\tt Y$, as the kernel interpolates linearly between the two limiting cases.
The identification of a common scenario for both transitions 
is assisted by recent reports in numerical simulations of yielding~\cite{AguirrePRE2018,JaglaPRE2017,ferrero-jagla}
about phenomenological properties akin to mean-field depinning.
In particular, dynamical critical exponents are seen to depend on the details
of the local disorder potential, a result that has long been known for mean-field
depinning~\cite{FisherPR1998,KoltonPRE2018}.

In recent years, different attempts have been made to address's
the yielding transition of amorphous solids from an analytical perspective.
Nevertheless, the noncompliance with the hypothesis needed for the 
Functional Renormalization Group analysis has largely confined these treatments to
mean-field H\'ebraud-Lequeux-like approaches~\cite{AgoritsasEPJE2015, JaglaPRE2015, AgoritsasSM2016}
and heavy-tail noise variants \cite{LinPRX2016, lin2017some, AguirrePRE2018, JaglaJSTAT2018}.
Those studies provided a common general picture but did not forge 
a consensus about critical exponents and scaling laws.
One of the main conclusions of the present work is that the yielding transition,
as described by a scalar model with an Eshelby interaction, can be treated
as a special case of a generalized mean-field problem which has the very
well known fully-connected MF-depinning problem as one limiting case.
Our work suggest therefore that, instead of focussing on non-trivial correlations depending
on the propagator properties and the dimension, an strategic angle of attack for theoretical
studies of the yielding transition could be to start from a fully-connected depinning system
and explore perturbations of the Eshelby type to the constant propagator.

{\it Acknowledgments--}
EEF acknowledges support from grant PICT 2017-1202, ANPCyT (Argentina).

%

\appendix

\subsection{Simulation protocols}\label{ap:protocol}

We do straightforward simulations of two-dimensional systems described
by Eq.(\ref{model}) integrated with a first order Euler method.
In each integration step, the term  $\sum_j G_{ij}e_j$ is treated in Fourier space,
computed as $\sum_{\bf q} G_{\bf q}e_{\bf q}$. 
In this respect, note that in a square numerical mesh of size $L\times L$,
quantities such as $q_x^2$ and $q_y^2$ must be understood  as 
\begin{equation}
q_{x,y}^2\equiv 2-2\cos\left (\frac{\pi n_{x,y}}{L}\right )
\end{equation}
with $n_{x,y}=0,..., L-1$.

Concerning the form of the local forces $f_i=-dV_i/e_i$, we generate them `on the flight'
in the following way.
We start with the value of $e_i$ such that $e_L<e_i<e_R$, where a 
parabolic potential is defined in term of $e_L$, $e_R$ and having an unitary curvature.
In concrete, the force on $e_i$ coming from the disordered potential  is taken as
$f_i=-(e_i-(e_L+e_R)/2)$.
As soon as the  dynamics makes $e_i$ larger than $e_R$, we set a new parabolic
potential well for $e_i$ by choosing
\begin{eqnarray}
e_L^{new}&=&e_R\\
e_R^{new}&=&e_R+\Delta
\end{eqnarray}
where the $\Delta$ is randomly chosen from a flat distribution between 0.5 and 1.5.
This is what we call the `cuspy' potential, since its composed by a concatenation of parabolic pieces
and the transition from one to another produces a discontinuity in the force.
In the case of `smooth potentials' instead, the potential wells are defined and updated in the same form,
but the force on each
potential well is given by
\begin{equation}
f_i=-\sin (e_i-(e_L+e_R)/2).
\end{equation}
Note that in this form, the value of $f_i$ and also its derivative $df_i/e_i$ are continuous functions of $e_i$.

In constant stress simulations, the value of $\sigma$ in Eq.(\ref{model}) is kept fixed,
and the main output of the simulation is the value of $\dot \gamma$, Eq.(\ref{gammadot}).
Such a protocol is used primarily to obtain the flow curves.
A second important outcome of these simulations, is the distribution of local 
distances to instability, namely the distribution of the quantities $x_i \equiv e_{Ri}-e_i$.
The average minimum value of $x_i$ is used to calculate the $\theta$ exponent. 

\begin{figure}[b]
\includegraphics[width=8cm,clip=true]{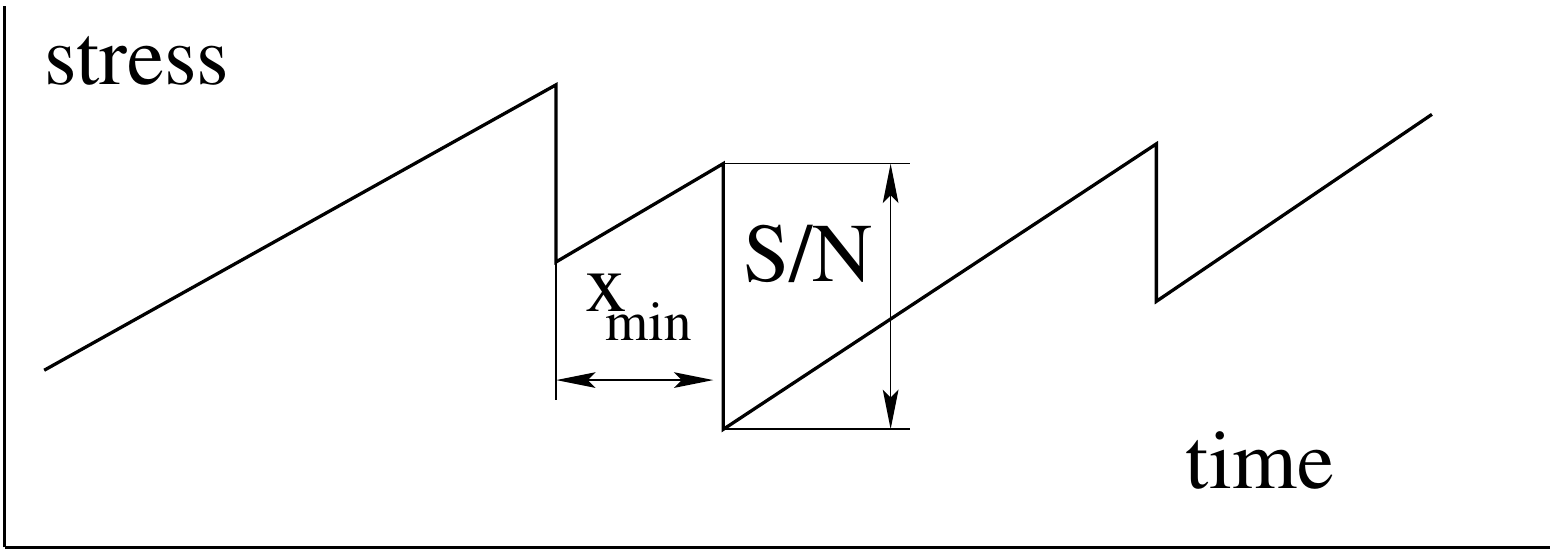}
\caption{\label{serrucho}
Sketch of the stress in the system as a function of time in quasistatic simulations.
From this kind of plot, statistics of avalanche size ($S$) and stress increase to destabilize
a new avalanche ($x_{\tt min}$) can be collected ($N\equiv L^d$ is the system size).
}
\end{figure}

On the other hand, we want to be able to access individual avalanches very close to the critical stress,
and collect statistics of size, duration, etc.
This can be better accomplished by using of a quasi-static protocol.
To do so, we move away from the fix-stress modelling and modify $G_{\bf q}$
by defining $G_{{\bf q}=0}=-\kappa$, with $\kappa$ a constant parameter of order one.
The equation for the evolution of the average strain is transformed to 
\begin{equation}
\dot\gamma\equiv \frac{d\overline {e_i}}{dt}=\overline{f_i(e_i)}- \overline {e_i}\kappa+ \sigma
\label{gamap}
\end{equation} 
that can be interpreted as a progressive reduction of $\sigma$ (due to the term $\overline {e_i}\kappa$)
as the average position of the interface moves forward.
This stress reduction guarantees that any activity in the system will eventually stop,
reaching a metastable static configuration.
At this point the stress has to be increased again to trigger a new avalanche, and the
process can be repeated.
The evolution of stress along the simulation is sketched in Fig. \ref{serrucho}.
From such a simulation we collect the statistics of avalanche size ($S$) and 
stress increments needed to trigger new avalanches $x_{\tt min}$.

It should be mentioned, nevertheless, that quasistatic simulations in the form just described
are rather inefficient:
The dynamic evolution is continuous and we need to wait until the activity falls below
a very low threshold to safely decide that the avalanche has stopped.
In the same way, to trigger a new avalanche, the stress has to be increased very slowly
to be sure to detect the precise beginning of the new avalanche.
In the case of piece-wise parabolic potentials, an accelerated numerical scheme can be implemented as follows.
Since we use a potential where all parabolic wells have the same curvature, the form of $f_i(e_i)$ is simply given by
\begin{equation}
f_i(e_i)=-(e_i-e_{0i})
\end{equation} 
with $e_{0i}=(e_{Li}+e_{Ri})/2$. 
Then Eq.(\ref{model}) can be solved in a single step to obtain the new equilibrium position of the interface.
The solution in Fourier space is given by
\begin{equation}
e_{\bf q}=\frac {e_{0\bf q}}{1-G_{\bf q}}.
\label{solve}
\end{equation} 
If all the $e_i$ obtained by Fourier-inverting (\ref{solve}) lay within their potential well,
i.e.,  $e_{Li}<e_i<e_{Ri}$, then the configuration found is a static solution to the problem.
However, if some resulting $e_i$ happen to be outside the range $\left[e_{Li},e_{Ri}\right]$,
it means that the corresponding $e_{0i}$ have to be adjusted and Eq.(\ref{solve}) solved again
to find a new set of $e_i$.
This process is repeated until all $e_i$ are within $\left[e_{Li},e_{Ri}\right]$.
At this point the avalanche stops.
In this scheme one actually loose the true continuous time evolution of the real dynamics, but it results 
to be computationally much more efficient and--as verified in test cases--it does not show
any noticeable differences in the avalanche statistics with respect to the case in which the true dynamics is used.


\begin{thebibliography}{25}%
\makeatletter
\providecommand \@ifxundefined [1]{%
 \@ifx{#1\undefined}
}%
\providecommand \@ifnum [1]{%
 \ifnum #1\expandafter \@firstoftwo
 \else \expandafter \@secondoftwo
 \fi
}%
\providecommand \@ifx [1]{%
 \ifx #1\expandafter \@firstoftwo
 \else \expandafter \@secondoftwo
 \fi
}%
\providecommand \natexlab [1]{#1}%
\providecommand \enquote  [1]{``#1''}%
\providecommand \bibnamefont  [1]{#1}%
\providecommand \bibfnamefont [1]{#1}%
\providecommand \citenamefont [1]{#1}%
\providecommand \href@noop [0]{\@secondoftwo}%
\providecommand \href [0]{\begingroup \@sanitize@url \@href}%
\providecommand \@href[1]{\@@startlink{#1}\@@href}%
\providecommand \@@href[1]{\endgroup#1\@@endlink}%
\providecommand \@sanitize@url [0]{\catcode `\\12\catcode `\$12\catcode
  `\&12\catcode `\#12\catcode `\^12\catcode `\_12\catcode `\%12\relax}%
\providecommand \@@startlink[1]{}%
\providecommand \@@endlink[0]{}%
\providecommand \url  [0]{\begingroup\@sanitize@url \@url }%
\providecommand \@url [1]{\endgroup\@href {#1}{\urlprefix }}%
\providecommand \urlprefix  [0]{URL }%
\providecommand \Eprint [0]{\href }%
\providecommand \doibase [0]{http://dx.doi.org/}%
\providecommand \selectlanguage [0]{\@gobble}%
\providecommand \bibinfo  [0]{\@secondoftwo}%
\providecommand \bibfield  [0]{\@secondoftwo}%
\providecommand \translation [1]{[#1]}%
\providecommand \BibitemOpen [0]{}%
\providecommand \bibitemStop [0]{}%
\providecommand \bibitemNoStop [0]{.\EOS\space}%
\providecommand \EOS [0]{\spacefactor3000\relax}%
\providecommand \BibitemShut  [1]{\csname bibitem#1\endcsname}%
\let\auto@bib@innerbib\@empty
\bibitem [{\citenamefont {Fisher}(1998)}]{FisherPR1998}%
  \BibitemOpen
  \bibfield  {author} {\bibinfo {author} {\bibfnamefont {D.~S.}\ \bibnamefont
  {Fisher}},\ }\href {https://doi.org/10.1016/S0370-1573(98)00008-8} {\bibfield
   {journal} {\bibinfo  {journal} {Phys. Rep.}\ }\textbf {\bibinfo {volume}
  {301}},\ \bibinfo {pages} {113} (\bibinfo {year} {1998})}\BibitemShut
  {NoStop}%
\bibitem [{\citenamefont {Kardar}(1998)}]{KardarPR1998}%
  \BibitemOpen
  \bibfield  {author} {\bibinfo {author} {\bibfnamefont {M.}~\bibnamefont
  {Kardar}},\ }\href {\doibase https://doi.org/10.1016/S0370-1573(98)00007-6}
  {\bibfield  {journal} {\bibinfo  {journal} {Physics Reports}\ }\textbf
  {\bibinfo {volume} {301}},\ \bibinfo {pages} {85 } (\bibinfo {year}
  {1998})}\BibitemShut {NoStop}%
\bibitem [{\citenamefont {Nicolas}\ \emph {et~al.}(2018)\citenamefont
  {Nicolas}, \citenamefont {Ferrero}, \citenamefont {Martens},\ and\
  \citenamefont {Barrat}}]{NicolasRMP2018}%
  \BibitemOpen
  \bibfield  {author} {\bibinfo {author} {\bibfnamefont {A.}~\bibnamefont
  {Nicolas}}, \bibinfo {author} {\bibfnamefont {E.~E.}\ \bibnamefont
  {Ferrero}}, \bibinfo {author} {\bibfnamefont {K.}~\bibnamefont {Martens}}, \
  and\ \bibinfo {author} {\bibfnamefont {J.-L.}\ \bibnamefont {Barrat}},\
  }\href {\doibase 10.1103/RevModPhys.90.045006} {\bibfield  {journal}
  {\bibinfo  {journal} {Rev. Mod. Phys.}\ }\textbf {\bibinfo {volume} {90}},\
  \bibinfo {pages} {045006} (\bibinfo {year} {2018})}\BibitemShut {NoStop}%
\bibitem [{\citenamefont {Talamali}\ \emph {et~al.}(2011)\citenamefont
  {Talamali}, \citenamefont {Pet\"{a}j\"{a}}, \citenamefont {Vandembroucq},\
  and\ \citenamefont {Roux}}]{TalamaliPRE2011}%
  \BibitemOpen
  \bibfield  {author} {\bibinfo {author} {\bibfnamefont {M.}~\bibnamefont
  {Talamali}}, \bibinfo {author} {\bibfnamefont {V.}~\bibnamefont
  {Pet\"{a}j\"{a}}}, \bibinfo {author} {\bibfnamefont {D.}~\bibnamefont
  {Vandembroucq}}, \ and\ \bibinfo {author} {\bibfnamefont {S.}~\bibnamefont
  {Roux}},\ }\href {https://doi.org/10.1103/PhysRevE.84.016115} {\bibfield
  {journal} {\bibinfo  {journal} {Physical Review E}\ }\textbf {\bibinfo
  {volume} {84}} (\bibinfo {year} {2011})}\BibitemShut {NoStop}%
\bibitem [{\citenamefont {Lin}\ \emph {et~al.}(2014{\natexlab{a}})\citenamefont
  {Lin}, \citenamefont {Lerner}, \citenamefont {Rosso},\ and\ \citenamefont
  {Wyart}}]{LinPNAS2014}%
  \BibitemOpen
  \bibfield  {author} {\bibinfo {author} {\bibfnamefont {J.}~\bibnamefont
  {Lin}}, \bibinfo {author} {\bibfnamefont {E.}~\bibnamefont {Lerner}},
  \bibinfo {author} {\bibfnamefont {A.}~\bibnamefont {Rosso}}, \ and\ \bibinfo
  {author} {\bibfnamefont {M.}~\bibnamefont {Wyart}},\ }\href
  {https://doi.org/10.1073/pnas.1406391111} {\bibfield  {journal} {\bibinfo
  {journal} {Proceedings of the National Academy of Sciences}\ }\textbf
  {\bibinfo {volume} {111}},\ \bibinfo {pages} {14382} (\bibinfo {year}
  {2014}{\natexlab{a}})}\BibitemShut {NoStop}%
\bibitem [{\citenamefont {Budrikis}\ \emph {et~al.}(2017)\citenamefont
  {Budrikis}, \citenamefont {Castellanos}, \citenamefont {Sandfeld},
  \citenamefont {Zaiser},\ and\ \citenamefont {Zapperi}}]{BudrikisNC2015}%
  \BibitemOpen
  \bibfield  {author} {\bibinfo {author} {\bibfnamefont {Z.}~\bibnamefont
  {Budrikis}}, \bibinfo {author} {\bibfnamefont {D.~F.}\ \bibnamefont
  {Castellanos}}, \bibinfo {author} {\bibfnamefont {S.}~\bibnamefont
  {Sandfeld}}, \bibinfo {author} {\bibfnamefont {M.}~\bibnamefont {Zaiser}}, \
  and\ \bibinfo {author} {\bibfnamefont {S.}~\bibnamefont {Zapperi}},\ }\href
  {https://www.nature.com/articles/ncomms15928} {\bibfield  {journal} {\bibinfo
   {journal} {Nat. Comm.}\ }\textbf {\bibinfo {volume} {8}},\ \bibinfo {pages}
  {15928} (\bibinfo {year} {2017})}\BibitemShut {NoStop}%
\bibitem [{\citenamefont {H\'{e}braud}\ and\ \citenamefont
  {Lequeux}(1998)}]{HebraudPRL1998}%
  \BibitemOpen
  \bibfield  {author} {\bibinfo {author} {\bibfnamefont {P.}~\bibnamefont
  {H\'{e}braud}}\ and\ \bibinfo {author} {\bibfnamefont {F.}~\bibnamefont
  {Lequeux}},\ }\href {https://doi.org/10.1103/PhysRevLett.81.2934} {\bibfield
  {journal} {\bibinfo  {journal} {Physical Review Letters}\ }\textbf {\bibinfo
  {volume} {81}},\ \bibinfo {pages} {2934} (\bibinfo {year}
  {1998})}\BibitemShut {NoStop}%
\bibitem [{\citenamefont {Agoritsas}\ \emph {et~al.}(2015)\citenamefont
  {Agoritsas}, \citenamefont {Bertin}, \citenamefont {Martens},\ and\
  \citenamefont {Barrat}}]{AgoritsasEPJE2015}%
  \BibitemOpen
  \bibfield  {author} {\bibinfo {author} {\bibfnamefont {E.}~\bibnamefont
  {Agoritsas}}, \bibinfo {author} {\bibfnamefont {E.}~\bibnamefont {Bertin}},
  \bibinfo {author} {\bibfnamefont {K.}~\bibnamefont {Martens}}, \ and\
  \bibinfo {author} {\bibfnamefont {J.-L.}\ \bibnamefont {Barrat}},\ }\href
  {\doibase 10.1140/epje/i2015-15071-x} {\bibfield  {journal} {\bibinfo
  {journal} {Eur. Phys. J. E}\ }\textbf {\bibinfo {volume} {38}},\ \bibinfo
  {pages} {71} (\bibinfo {year} {2015})}\BibitemShut {NoStop}%
\bibitem [{\citenamefont {Jagla}(2015)}]{JaglaPRE2015}%
  \BibitemOpen
  \bibfield  {author} {\bibinfo {author} {\bibfnamefont {E.~A.}\ \bibnamefont
  {Jagla}},\ }\href {\doibase 10.1103/PhysRevE.92.042135} {\bibfield  {journal}
  {\bibinfo  {journal} {Phys. Rev. E}\ }\textbf {\bibinfo {volume} {92}},\
  \bibinfo {pages} {042135} (\bibinfo {year} {2015})}\BibitemShut {NoStop}%
\bibitem [{\citenamefont {Agoritsas}\ and\ \citenamefont
  {Martens}(2017)}]{AgoritsasSM2016}%
  \BibitemOpen
  \bibfield  {author} {\bibinfo {author} {\bibfnamefont {E.}~\bibnamefont
  {Agoritsas}}\ and\ \bibinfo {author} {\bibfnamefont {K.}~\bibnamefont
  {Martens}},\ }\href {\doibase 10.1039/C6SM02702D} {\bibfield  {journal}
  {\bibinfo  {journal} {Soft Matter}\ }\textbf {\bibinfo {volume} {13}},\
  \bibinfo {pages} {4653} (\bibinfo {year} {2017})}\BibitemShut {NoStop}%
\bibitem [{\citenamefont {Lin}\ and\ \citenamefont {Wyart}(2016)}]{LinPRX2016}%
  \BibitemOpen
  \bibfield  {author} {\bibinfo {author} {\bibfnamefont {J.}~\bibnamefont
  {Lin}}\ and\ \bibinfo {author} {\bibfnamefont {M.}~\bibnamefont {Wyart}},\
  }\href {https://doi.org/10.1103/PhysRevX.6.011005} {\bibfield  {journal}
  {\bibinfo  {journal} {Physical Review X}\ }\textbf {\bibinfo {volume} {6}},\
  \bibinfo {pages} {011005} (\bibinfo {year} {2016})}\BibitemShut {NoStop}%
\bibitem [{\citenamefont {Lin}\ and\ \citenamefont
  {Wyart}(2018)}]{lin2017some}%
  \BibitemOpen
  \bibfield  {author} {\bibinfo {author} {\bibfnamefont {J.}~\bibnamefont
  {Lin}}\ and\ \bibinfo {author} {\bibfnamefont {M.}~\bibnamefont {Wyart}},\
  }\href {\doibase 10.1103/PhysRevE.97.012603} {\bibfield  {journal} {\bibinfo
  {journal} {Phys. Rev. E}\ }\textbf {\bibinfo {volume} {97}},\ \bibinfo
  {pages} {012603} (\bibinfo {year} {2018})}\BibitemShut {NoStop}%
\bibitem [{\citenamefont {Fern\'andez~Aguirre}\ and\ \citenamefont
  {Jagla}(2018)}]{AguirrePRE2018}%
  \BibitemOpen
  \bibfield  {author} {\bibinfo {author} {\bibfnamefont {I.}~\bibnamefont
  {Fern\'andez~Aguirre}}\ and\ \bibinfo {author} {\bibfnamefont {E.~A.}\
  \bibnamefont {Jagla}},\ }\href {\doibase 10.1103/PhysRevE.98.013002}
  {\bibfield  {journal} {\bibinfo  {journal} {Phys. Rev. E}\ }\textbf {\bibinfo
  {volume} {98}},\ \bibinfo {pages} {013002} (\bibinfo {year}
  {2018})}\BibitemShut {NoStop}%
\bibitem [{\citenamefont {Jagla}(2018)}]{JaglaJSTAT2018}%
  \BibitemOpen
  \bibfield  {author} {\bibinfo {author} {\bibfnamefont {E.~A.}\ \bibnamefont
  {Jagla}},\ }\href {\doibase 10.1088/1742-5468/aa9db2} {\bibfield  {journal}
  {\bibinfo  {journal} {Journal of Statistical Mechanics: Theory and
  Experiment}\ }\textbf {\bibinfo {volume} {2018}},\ \bibinfo {pages} {013401}
  (\bibinfo {year} {2018})}\BibitemShut {NoStop}%
\bibitem [{\citenamefont {Eshelby}(1957)}]{Eshelby1957}%
  \BibitemOpen
  \bibfield  {author} {\bibinfo {author} {\bibfnamefont {J.}~\bibnamefont
  {Eshelby}},\ }\href {https://doi.org/10.1098/rspa.1957.0133} {\bibfield
  {journal} {\bibinfo  {journal} {Proceedings of the Royal Society A:
  Mathematical, Physical and Engineering Sciences}\ }\textbf {\bibinfo {volume}
  {241}},\ \bibinfo {pages} {376} (\bibinfo {year} {1957})}\BibitemShut
  {NoStop}%
\bibitem [{\citenamefont {Picard}\ \emph {et~al.}(2004)\citenamefont {Picard},
  \citenamefont {Ajdari}, \citenamefont {Lequeux},\ and\ \citenamefont
  {Bocquet}}]{Picard2004}%
  \BibitemOpen
  \bibfield  {author} {\bibinfo {author} {\bibfnamefont {G.}~\bibnamefont
  {Picard}}, \bibinfo {author} {\bibfnamefont {A.}~\bibnamefont {Ajdari}},
  \bibinfo {author} {\bibfnamefont {F.}~\bibnamefont {Lequeux}}, \ and\
  \bibinfo {author} {\bibfnamefont {L.}~\bibnamefont {Bocquet}},\ }\href
  {https://doi.org/10.1140/epje/i2004-10054-8} {\bibfield  {journal} {\bibinfo
  {journal} {The European physical journal. E, Soft matter}\ }\textbf {\bibinfo
  {volume} {15}},\ \bibinfo {pages} {371} (\bibinfo {year} {2004})}\BibitemShut
  {NoStop}%
\bibitem [{\citenamefont {Kolton}\ and\ \citenamefont
  {Jagla}(2018)}]{KoltonPRE2018}%
  \BibitemOpen
  \bibfield  {author} {\bibinfo {author} {\bibfnamefont {A.~B.}\ \bibnamefont
  {Kolton}}\ and\ \bibinfo {author} {\bibfnamefont {E.~A.}\ \bibnamefont
  {Jagla}},\ }\href {\doibase 10.1103/PhysRevE.98.042111} {\bibfield  {journal}
  {\bibinfo  {journal} {Phys. Rev. E}\ }\textbf {\bibinfo {volume} {98}},\
  \bibinfo {pages} {042111} (\bibinfo {year} {2018})}\BibitemShut {NoStop}%
\bibitem [{Note1()}]{Note1}%
  \BibitemOpen
  \bibinfo {note} {$\sigma _c$ depends on the value of $\varepsilon $ and must
  be fitted for each curve}\BibitemShut {NoStop}%
\bibitem [{Note2()}]{Note2}%
  \BibitemOpen
  \bibinfo {note} {For a better comparison, in Fig.~\ref {avalanches} the
  values of $\kappa $ were adjusted to obtain curves with a similar
  cutoff.}\BibitemShut {Stop}%
\bibitem [{\citenamefont {Maloney}\ and\ \citenamefont
  {Lema\^{\i}tre}(2004)}]{Maloney2004}%
  \BibitemOpen
  \bibfield  {author} {\bibinfo {author} {\bibfnamefont {C.}~\bibnamefont
  {Maloney}}\ and\ \bibinfo {author} {\bibfnamefont {A.}~\bibnamefont
  {Lema\^{\i}tre}},\ }\href {https://doi.org/10.1103/PhysRevLett.93.016001}
  {\bibfield  {journal} {\bibinfo  {journal} {Physical Review Letters}\
  }\textbf {\bibinfo {volume} {93}},\ \bibinfo {pages} {016001} (\bibinfo
  {year} {2004})}\BibitemShut {NoStop}%
\bibitem [{\citenamefont {Karmakar}\ \emph {et~al.}(2010)\citenamefont
  {Karmakar}, \citenamefont {Lerner},\ and\ \citenamefont
  {Procaccia}}]{KarmakarPRE2010}%
  \BibitemOpen
  \bibfield  {author} {\bibinfo {author} {\bibfnamefont {S.}~\bibnamefont
  {Karmakar}}, \bibinfo {author} {\bibfnamefont {E.}~\bibnamefont {Lerner}}, \
  and\ \bibinfo {author} {\bibfnamefont {I.}~\bibnamefont {Procaccia}},\ }\href
  {\doibase 10.1103/PhysRevE.82.055103} {\bibfield  {journal} {\bibinfo
  {journal} {Phys. Rev. E}\ }\textbf {\bibinfo {volume} {82}},\ \bibinfo
  {pages} {055103} (\bibinfo {year} {2010})}\BibitemShut {NoStop}%
\bibitem [{\citenamefont {Lin}\ \emph {et~al.}(2014{\natexlab{b}})\citenamefont
  {Lin}, \citenamefont {Saade}, \citenamefont {Lerner}, \citenamefont {Rosso},\
  and\ \citenamefont {Wyart}}]{LinEPL2014}%
  \BibitemOpen
  \bibfield  {author} {\bibinfo {author} {\bibfnamefont {J.}~\bibnamefont
  {Lin}}, \bibinfo {author} {\bibfnamefont {A.}~\bibnamefont {Saade}}, \bibinfo
  {author} {\bibfnamefont {E.}~\bibnamefont {Lerner}}, \bibinfo {author}
  {\bibfnamefont {A.}~\bibnamefont {Rosso}}, \ and\ \bibinfo {author}
  {\bibfnamefont {M.}~\bibnamefont {Wyart}},\ }\href
  {https://doi.org/10.1209/0295-5075/105/26003} {\bibfield  {journal} {\bibinfo
   {journal} {Europhysics Letters (EPL)}\ }\textbf {\bibinfo {volume} {105}},\
  \bibinfo {pages} {26003} (\bibinfo {year} {2014}{\natexlab{b}})}\BibitemShut
  {NoStop}%
\bibitem [{Note3()}]{Note3}%
  \BibitemOpen
  \bibinfo {note} {In a driving protocol towards the right we define
  $x_i=e_{Ri}-e_i$ according to the notation in Appendix \ref
  {ap:protocol}}\BibitemShut {NoStop}%
\bibitem [{\citenamefont {Jagla}(2017)}]{JaglaPRE2017}%
  \BibitemOpen
  \bibfield  {author} {\bibinfo {author} {\bibfnamefont {E.~A.}\ \bibnamefont
  {Jagla}},\ }\href {\doibase 10.1103/PhysRevE.96.023006} {\bibfield  {journal}
  {\bibinfo  {journal} {Phys. Rev. E}\ }\textbf {\bibinfo {volume} {96}},\
  \bibinfo {pages} {023006} (\bibinfo {year} {2017})}\BibitemShut {NoStop}%
\bibitem [{\citenamefont {Ferrero}\ and\ \citenamefont
  {Jagla}()}]{ferrero-jagla}%
  \BibitemOpen
  \bibfield  {author} {\bibinfo {author} {\bibfnamefont {E.~E.}\ \bibnamefont
  {Ferrero}}\ and\ \bibinfo {author} {\bibfnamefont {E.~A.}\ \bibnamefont
  {Jagla}},\ }\href {https://arxiv.org/abs/1905.05610} {\enquote {\bibinfo
  {title} {Static and dynamic critical exponents in elastoplastic models of
  amorphous solids},}\ }\bibinfo {howpublished} {preprint -
  arXiv:1905.05610}\BibitemShut {NoStop}%
\end{thebibliography}
\end{document}